\documentclass[aps,pre,twocolumn,superscriptaddress,floatfix,showpacs]{revtex4}

\usepackage{amssymb}
\usepackage{amsmath}
\usepackage{latexsym}
\usepackage{graphicx}
\usepackage{subfigure}

\begin{document}

\title{Optimization in random field Ising models by quantum annealing}

\author{Matti Sarjala}
\affiliation{
Helsinki University of Techn., Lab. of Physics,
P.O.Box 1100, 02015 HUT, Finland}
\author{Viljo Pet\"aj\"a}
\affiliation{
Helsinki University of Techn., Lab. of Physics,
P.O.Box 1100, 02015 HUT, Finland}
\author{Mikko Alava}
\affiliation{
Helsinki University of Techn., Lab. of Physics,
P.O.Box 1100, 02015 HUT, Finland}

\newcommand{\bc}{\begin{center}}
\newcommand{\ec}{\end{center}}
\newcommand{\be}{\begin{equation}}
\newcommand{\ee}{\end{equation}}
\newcommand{\ba}{\begin{array}}
\newcommand{\ea}{\end{array}}
\newcommand{\beqn}{\begin{eqnarray}}
\newcommand{\eeqn}{\end{eqnarray}}

\begin{abstract}
We investigate the properties of quantum annealing applied to the
random field Ising model in one, two and three dimensions.  The decay
rate of the residual energy, defined as the energy excess from
the ground state, is find to be $e_{res}\sim
\log(N_{MC})^{-\zeta}$ with $\zeta$ in the range $2...6$, depending on
the strength of the random field. Systems with ``large clusters''
are harder to optimize as measured by $\zeta$. 
Our numerical results suggest that in the
ordered phase $\zeta=2$ whereas in the paramagnetic phase the
annealing procedure can be tuned so that $\zeta\to6$.
\end{abstract}


\maketitle

\section{Introduction}

It is desirable to have an optimization method which can be applied to
as wide range of problems. An example method is the classical
simulated annealing (SA).  During the recent years quantum annealing
(QA) has gained a lot of attention as a promising candidate for a
method, with a promise of a faster convergence to the optimal
configuration for a given problem.  This has been partly motivated by
real world realizations, demonstrated in experiments by Brooke et
al. \cite{brooke}.

Test problems for QA can be found from many problem--specific
optimization algorithms which find the exact ground state in a
polynomial time, as for instance the random field Ising model or the
Ising spin glass in two dimensions \cite{alava01}.  A convenient
measure for the effectiveness of the annealing is the residual energy
$e_{res}$ which gives the energy difference between the true ground
state and the configuration that is obtained in the end of the
annealing process.  The most important quantity is the annealing time
$\tau$, during which the temperature (transverse field in the case of
quantum annealing) of the system is reduced to zero. When $\tau$ is
increased the energy of the resulting configuration approaches the
ground state energy.
 
For classical simulated annealing it has been predicted by Huse
and Fisher \cite{huse} that the residual energy $e_{res}$ decreases with
the annealing time $\tau$ as
\be
e_{res}\sim\log(\tau)^{-\zeta} \quad. 
\label{e_vs_t}
\ee
For large time scales they have derived  
an upper limit for the decay of the residual energy in two-level
systems: $\zeta\le 2$.  This result is argued to hold also for random
field magnets and other disordered spin systems.  The existence of a
phase transition can change the value of $\zeta$.  According to Ref.\
\cite{huse} when a random field magnet is cooled through the phase
boundary from the disordered to the ordered phase the annealing slows
down to $\zeta=1$.

In Ref.\ \cite{santoro} Santoro et al. have studied quantum annealing
in the case of a two dimensional (2d) spin glass.  They have derived a
theoretical estimate for the decay rate of the residual
energy. Santoro et al. argued that the residual energy comes from
tunnelings at (avoided) Landau-Zener (LZ) crossings.  The
corresponding average residual energy resulting from this process was
estimated as
$e_{res}(\tau)=\int_0^{\Gamma_0} d\Gamma\, Z(\Gamma)E_{ex}(\Gamma)
\exp(\tau/\tau_c(\Gamma))$ 
where $Z(\Gamma)$ is the density of LZ crossings and $E_{ex}(\Gamma)$
is the corresponding average excitation energy.  The term
$\exp(\tau/\tau_c(\Gamma))$ gives the probability that the system
tunnels during the annealing to a higher energy eigen-state due to LZ
crossings, and $\tau_c(\Gamma)\sim\exp(A/\xi(\Gamma))$ where $\xi$ is
a typical wave localization length. Ref.\ \cite{santoro}
now considers the limit $\Gamma \to 0$ which is expected to dominate
the annealing behavior.  With the assumptions
$\xi(\Gamma)\sim\Gamma^{\varphi}$ and
$Z(\Gamma)E_{ex}(\Gamma)\sim\Gamma^{\omega}$
one gets $e_{res}\sim\log(\tau)^{-\zeta}$ with
$\zeta=(1+\omega)/\varphi$. With estimates $\varphi=1/2$ and $\omega=2$
given in Ref. \cite{santoro} one gets $\zeta=6$. 
The value for $\varphi$ comes from a quasi-classical consideration
of a particle's wave length, 
whereas $\omega=2$ can be reasoned for as follows.  In the limit
$\Gamma\to 0$ the tranverse field can be considered as a perturbation
for which only the second order correction to the energy has a
non-zero value. From this follows that $E_{ex}\sim\Gamma^2$. If the
density of LZ crossings in the limit $\Gamma\to 0$ is assumed to be at
most the density of the classical states at $\Gamma=0$ one gets
$\omega=2$.  Thus the residual energy decreases logarithmically but
now with much larger exponent $\zeta\approx 6$. This implies a
considerable speed-up compared to $\zeta=2$, the upper limit of SA. It
is useful to emphasize that this estimate for $\zeta$ is obtained
without any problem--specific assumptions.

The numerical results of Santoro et al. \cite{santoro} show that
the residual energy of the 2d Edwards-Anderson spin glass converges
indeed at much faster rate and to lower values with QA, compared to SA,
though no empirical values of $\zeta$ were given.
In addition, the performance of QA has been tested also on the 
traveling salesman problem \cite{TSP} for which similarly to 2d spin glass 
it was found that QA gives a faster decay of the residual energy than SA.  
However, this is not generally valid for all optimization problems. Battaglia
et al. \cite{3-sat} have found that in the case of the three-satisfiability
problem quantum annealing is outperformed by simulated annealing.
In small systems, it is possible to see a power law decay of  the
residual energy, e.g. by solving  the time-dependent 
Schr\"odinger equation adiabatically \cite{suzuki-okada,stella}. 
However, with increasing system sizes the energy gap  for
the Landau-Zener crossings decreases \cite{dziarmaga}, which means that the
probability of staying in the ground state decreases as well leading 
to the logarithmic behavior discussed above.

In this paper we study the quantum annealing applied to the random
field Ising model. The RFIM has the following Hamiltonian:
\be
H=-J\sum_{\langle i,j \rangle}s_is_j - \sum_i h_i s_i,
\label{h_cl}
\ee
where $J>0$ is the coupling constant, $s_i=\pm 1$ are classical spin
variables and $h_i$ is the random field at site $i$.
One of the advantages of RFIM as a test problem is the fact that its exact
ground state can be found in a polynomial time with an efficient graph
algorithm from combinatorial optimization \cite{alava01}. This allows
us to calculate the residual energy as the difference between
the true ground state and the configuration given by any annealing
procedure. 

Another feature of RFIM is the fact that the phase diagram
depends on both dimension and the strength of disorder. 
The second order phase transition of the 2d Ising model is
destroyed by the random fields, though residual ordering
persists in finite systems \cite{rfim_eira}.
Though there is thus no long range order either in one or two dimensions,
systems of a finite size may have ordered ground states when the
typical cluster size exceeds the system size, which is true also
at zero temperature \cite{rfim_eira,1drfim,rfim_2d}. However, in three
dimensions one has a temperature dependent critical strength of the
random field $h_c(T)$ below which the system is ordered
\cite{rfim_3d}. At zero temperature its value has been calculated
numerically as $h_c(T=0)=2.27$ \cite{middleton}.

We calculate the residual energy as a function of annealing time
measured in Monte Carlo steps in one, two and three dimensions with
varying strength of disorder.  Our numerical results suggest that the
residual energy decays logarithmically as in Eq.\ (\ref{e_vs_t}). However,
the value of the exponent $\zeta$ now seems to depend on the nature of
the ground state. In the ferromagnetic case $\zeta\approx 2$, whereas
in the paramagnetic case $\zeta\approx 6$, if the cooling schedule is
tuned well enough.  To our knowledge this is the first time when
empirical values for $\zeta$ are presented if not counting the simple
models giving a power law dacay of the residual energy
\cite{suzuki-okada,stella}.

The structure of the rest of the paper is the following.  In section
\ref{sec:numerics} we briefly review the Suzuki-Trotter mapping which
transforms a $d$-dimensional quantum system to a $d+1$-dimensional
classical one making the problem accessible for conventional Monte
Carlo sampling.  Section \ref{sec:results} is devoted to the numerical
results starting with the results for the classical simulated
annealing, which serve as a measuring stick when the efficiency of the
quantum annealing is discussed later on. For quantum annealing we
discuss numerical results in one, two and three dimensions for
different random field strengths.  It is also studied how the
performance of QA is affected when the parameters of the
$d+1$-dimensional equivalent classical system are varied. The paper is
summarized in Section \ref{sec:summary}.

\section{Numerics}\label{sec:numerics}

The quantum version of Eq.~(\ref{h_cl}) is obtained by replacing the spin
variables $s_i$ with Pauli spin operators $\sigma_i^z$.
Quantum fluctuations are tuned by changing the strength $\Gamma$
of a perpendicular field term, arising from the
Pauli spin operator $\sigma_i^x$. 
\be
H_{Q}=-J\sum_{\langle i,j \rangle}\sigma^z_i \sigma^z_j - \sum_i h_i
\sigma^z_i - \Gamma \sum_i \sigma^x_i.
\label{h_qt}
\ee
In the quantum annealing one starts with a large value of $\Gamma$ so
that spins in the $z$ direction are totally uncorrelated. By decreasing
$\Gamma$ gradually towards zero spins fall into the ground state
configuration provided that $T=0$.

With the Suzuki--Trotter mapping \cite{suzuki} this $d$-dimensional quantum
system can be represented by $P$ coupled replicas of the classical
system, Eq.~(\ref{h_cl}) resulting in a $d+1$ dimensional classical problem
with the following Hamiltonian:
\be
H_{ST}=-\sum_{k=1}^P \left( J\sum_{\langle i,j\rangle}s^k_i s^k_j 
+ \sum_i h_i s^k_i + J_{\perp}\sum_i s^k_i s^{k+1}_i \right),
\label{h_st}
\ee
where $J_{\perp}$ is the $\Gamma$ -dependent coupling constant between
the replicas:
\be
J_{\perp}=-\frac{PT}{2}\ln \mathrm{tanh}\frac{\Gamma}{PT}.
\label{j_perp}
\ee
The resulting system has periodic boundary conditions in the extra
dimension. It is convenient to set periodic boundaries also for the
original classical system.  The annealing of the Hamiltonian
(Eq.\ (\ref{h_qt})) is simulated by a standard Monte Carlo sampling of
Eq.\ (\ref{h_st}) at the effective temperature $PT$ with gradually
decreasing $\Gamma$.

The values of random field $h_i$ are taken from the fixed Gaussian
distribution $P_G(h_i)$ with the parameters $\langle h_i\rangle=0$ and
$\langle h_i^2\rangle=1$. The strength of the random field is tuned by
varying the ferromagnetic coupling constant $J$.  The residual energy
$e_{res}$ is calculated as
\be
e_{res}(N_{MC})=\langle E_{cl}(N_{MC})\rangle_k-E_{GS},
\label{e_res}
\ee
where $\langle E_{cl} \rangle_k$ is the average energy of all replicas
and $E_{GS}$ is the ground state energy. The GS energy and
configuration are computed for each sample, as noted in the
introduction, by using a combinatorial optimization algorithm.  Both
$\langle E_{cl}(N_{MC})\rangle_k$ and $E_{GS}$ from the definition in
Eq.\ (\ref{e_res}) are normalized per spin.

\section{Results}\label{sec:results}

\subsection{Classical simulated annealing}

\begin{figure}[t]
\includegraphics[width=0.85\columnwidth,angle=-90]{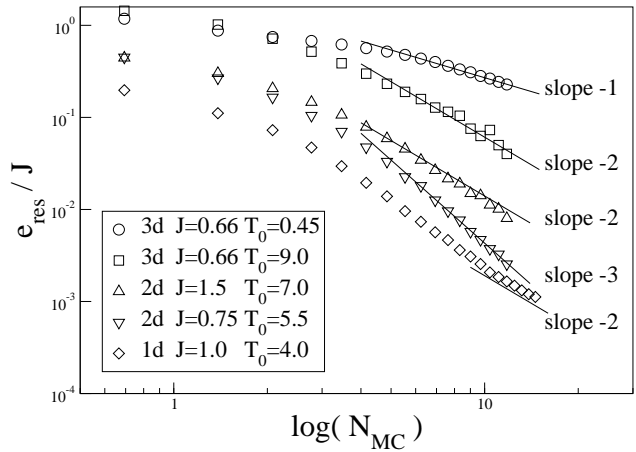}
\caption{\label{datSA} 
  A doubly logarithmic plot of residual energies in classical
  simulated annealing. $L=32$ in 3d, $L=256$, in 2d and $L=10^4$ in 1d
  with different coupling constants $J$ and initial temperatures
  $T_0$. The straight lines are guides for the eye.
}
\end{figure}

First, we briefly present numerical results for classical simulated
annealing, which are then later utilized in a comparison with quantum
annealing. Since the differences between various trial annealing
schedules in the case of SA turned out to be rather small
we show in Fig.\ \ref{datSA} only
the data that corresponds to a linear cooling schedule.  In one and
two dimensions we find the expected logarithmic decrease of the
residual energy ($e_{res}\sim\log(N_{MC})^{-\zeta}$) with
$\zeta=2...3$. According to Ref.\ \cite{huse} $\zeta=2$ is an upper
limit,  and hence it is expected that asymptotically $\zeta\to 2$.

In 3d in the ordered phase, for $J>J_c\approx 0.44$ \cite{middleton},
 one can observe the
slowing effect of the phase transition on the cooling efficiency in
agreement with the theoretically predicted $\zeta=1$ \cite{huse}. This is
clearly visible for the case where the annealing is started at low
temperature (ovals). When the starting temperature is well above $T_c$
the system stays in the paramagnetic phase during most of the simulation time
resulting in $\zeta=2$.  With increasing number of Monte Carlo steps the
system is expected to spend more annealing steps in the ferromagnetic
phase and hence to experience the slowing of the annealing rate to $\zeta=1$.

\subsection{Quantum annealing: preliminaries}

A priori it is naturally not clear which is the best way to reduce the value
of $\Gamma$. We have tested the three following annealing schedules:
\be
\Gamma_{\mathrm{Lin}}(N) = \Gamma_0 (1-\frac{N}{N_{MC}}) ,
\label{g_lin}
\ee
\be
\Gamma_{\mathrm R}(N) = \Gamma_0 / (1+R\frac{N}{N_{MC}})\quad , \quad 
R=\frac{\Gamma_0}{\Gamma_{f}}-1 ,
\label{g_r}
\ee
\beqn
\label{g_log}
\Gamma_{\mathrm{Log}}(N)&=&-\log \big\{\tanh\big[ \,\,\mathrm{atanh}(e^{-\Gamma_0})\quad\quad\quad\quad \\ 
& - &\frac{N}{N_{MC}}\left( \mathrm{atanh}(e^{-\Gamma_0})-\mathrm{atanh}(e^{-\Gamma_{f}}) \right)
  \big] \big\} . \nonumber
\eeqn

\begin{figure}[t]
\includegraphics[width=0.85\columnwidth,angle=-90]{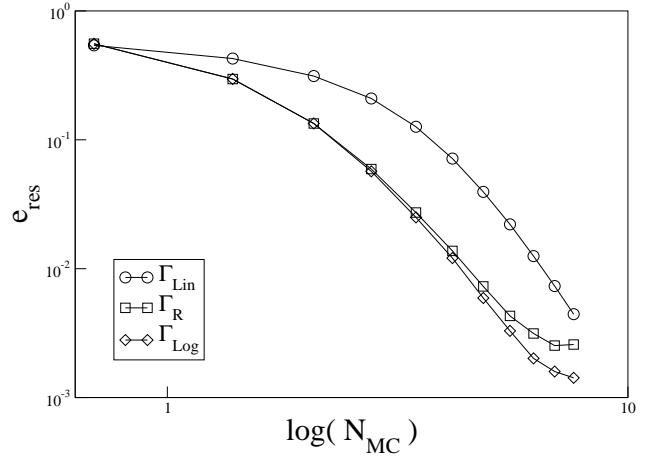}
\caption{\label{fig1} 
A doubly logarithmic plot of residual energies in 1d
corresponding to different quantum annealing schedules
Eqs. (\ref{g_lin})-(\ref{g_log}). 
}
\end{figure}

The residual energies corresponding to the different schedules
(Eqs.\ (\ref{g_lin})-(\ref{g_log})) for 1d systems of $10^4$
spins (averaged over more than 10 samples each) are shown in Fig.\
\ref{fig1}. With the used set of parameters ($J=1, PT=4, P=128,
\Gamma_0=8, \Gamma_f=10^{-6}$) the residual energies seem to decay
with the same slope for all annealing schedules. However, the
logarithmic schedule (Eq.\ (\ref{g_log})) gives the lowest residual energy
and hence we restrict ourselves to it throughout the rest of the paper. 
The choice of
$\Gamma_0$ does not alter the results as far as it is chosen large
enough in order to ensure that there are no correlations in the starting
configuration.  We flip only one spin at a time. We
tested also the use of the global flips where one attempts to flip all
replicas of a given spin. It turned out that the single flip
strategy is more effective.

\begin{figure}[t]\centering
\begin{minipage}{0.47\columnwidth}
\includegraphics[width=0.9\columnwidth]{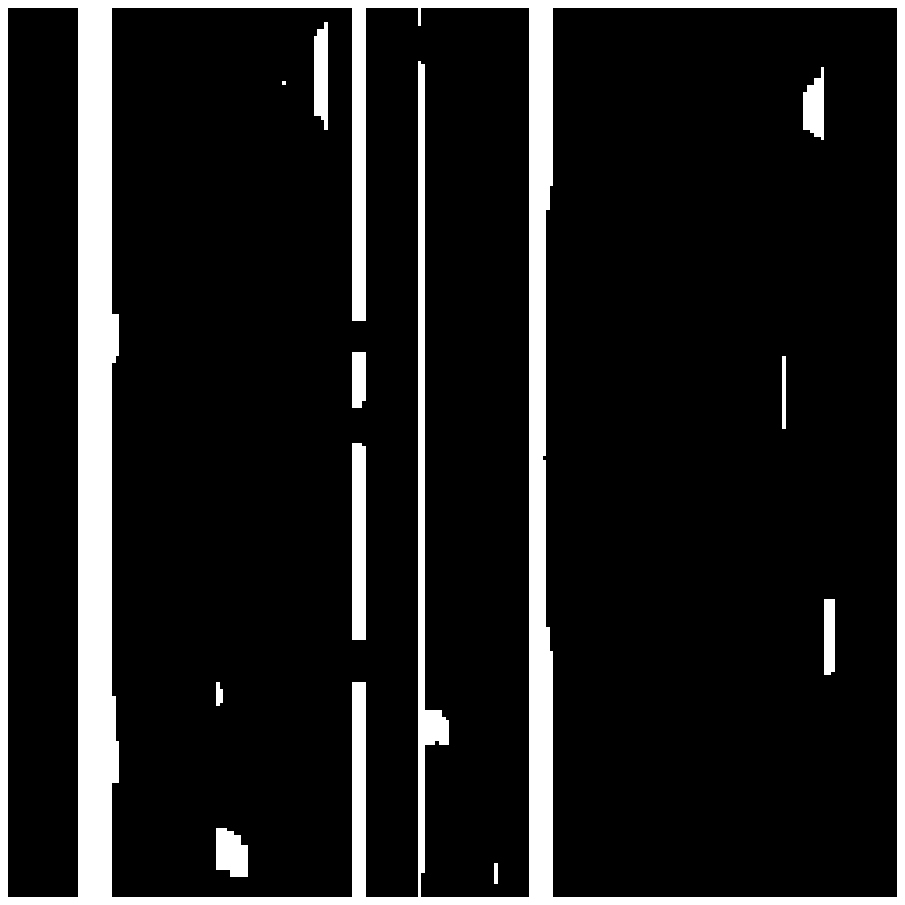}
\includegraphics[width=0.9\columnwidth]{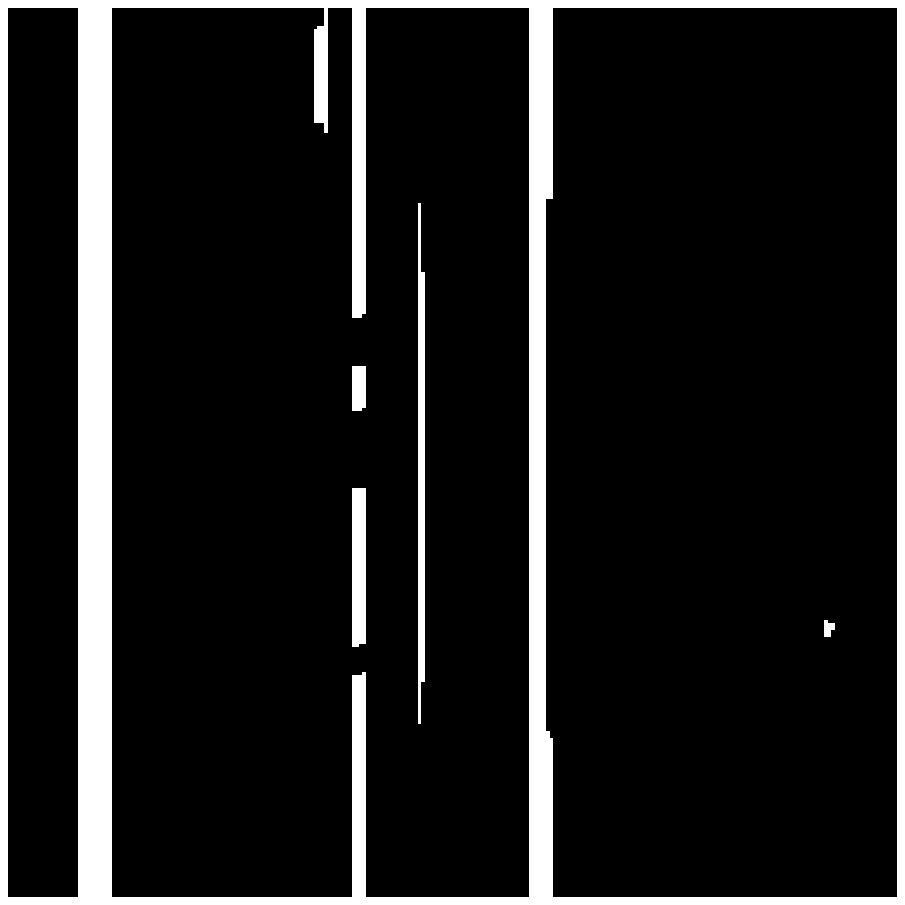}
\includegraphics[width=0.9\columnwidth]{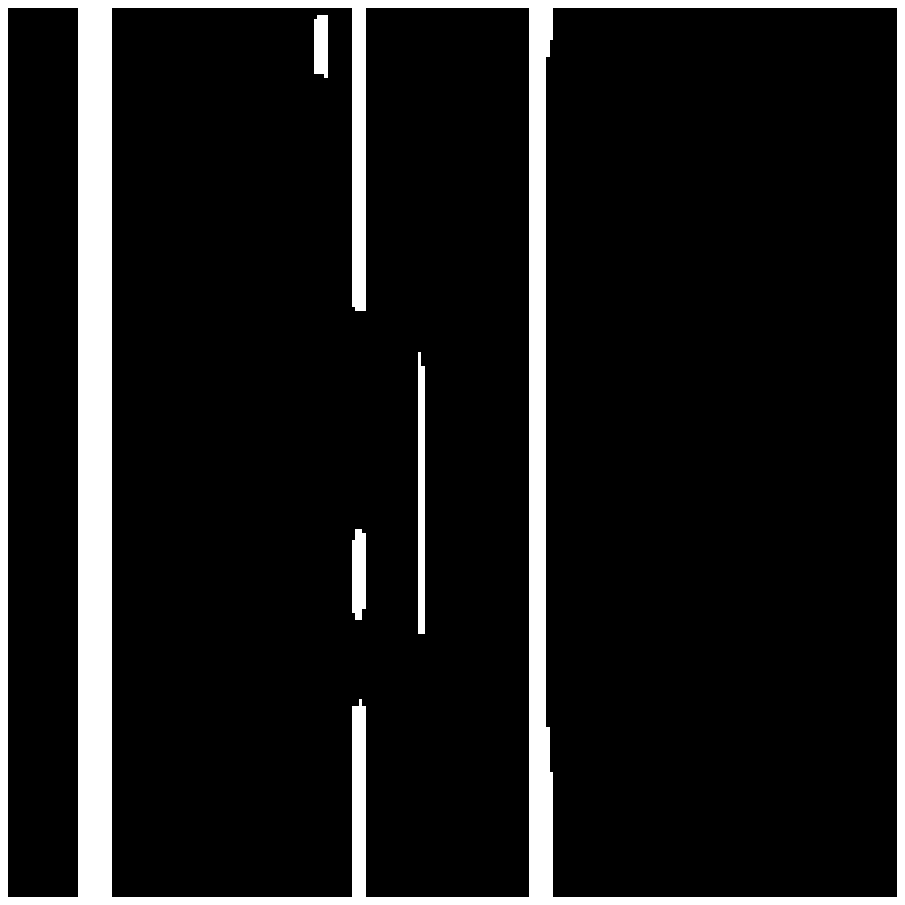}
\end{minipage}
\begin{minipage}{0.47\columnwidth}
\includegraphics[width=0.9\columnwidth]{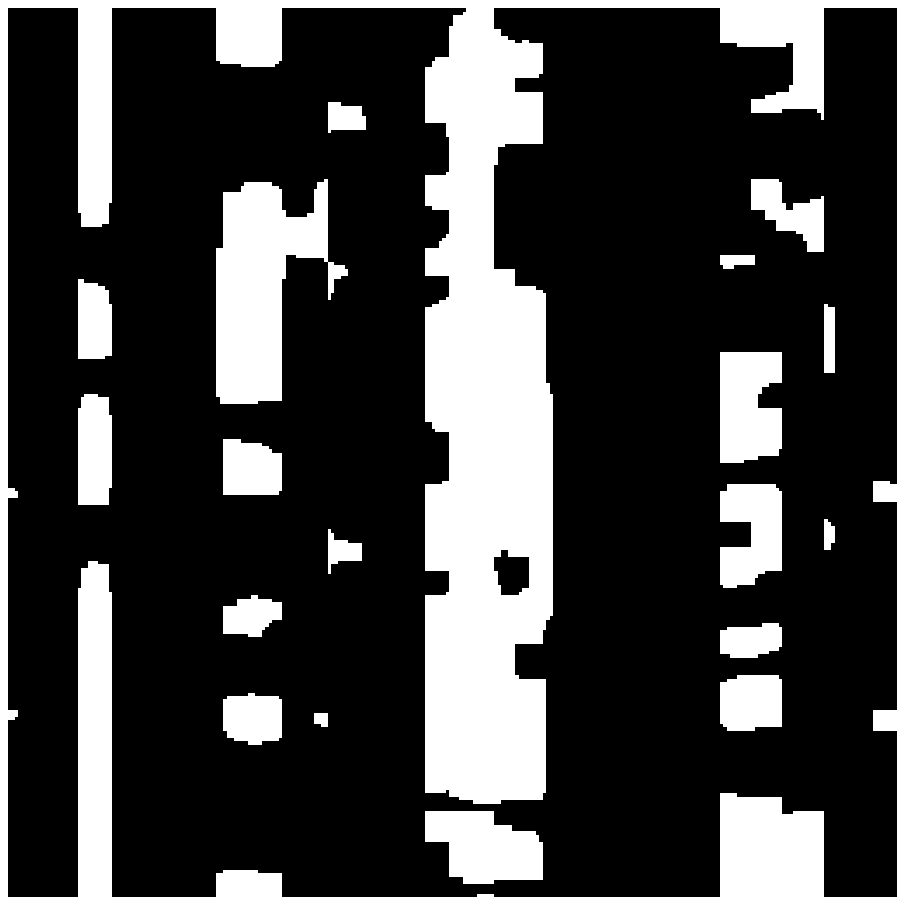}
\includegraphics[width=0.9\columnwidth]{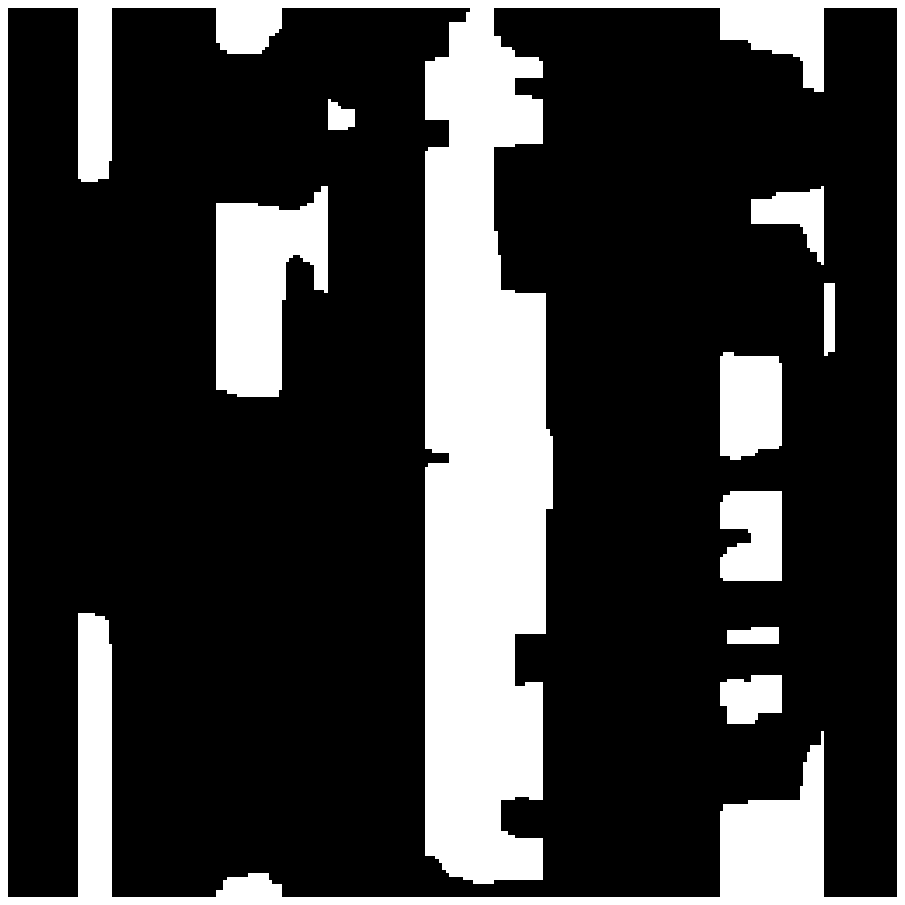}
\includegraphics[width=0.9\columnwidth]{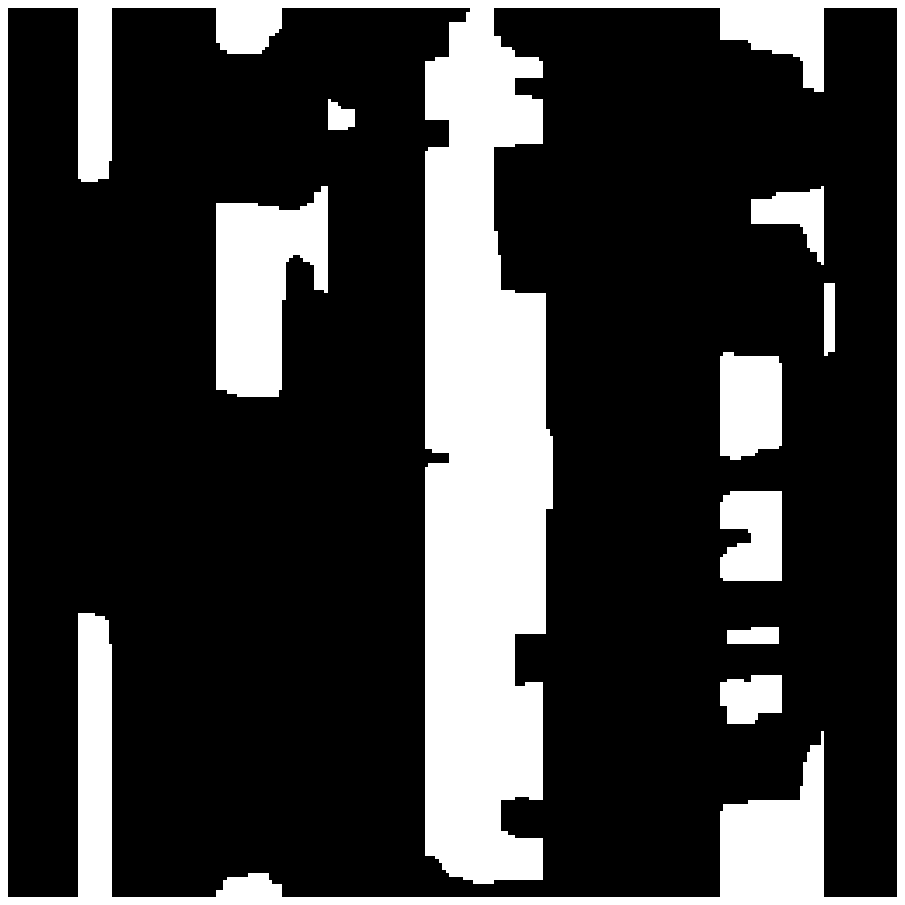}
\end{minipage}

\caption{\label{fig2} Snapshots of the spin configurations in the
  process of the annealing after (from top to bottom) $1/3 N_{MC}$, 
  $2/3 N_{MC}$ and $N_{MC}$ (final) Monte Carlo iterations. Black color
  denotes correctly aligned spins with respect to the ground
  state. Right: $J=2.0$, left: $J=4.0$ ($L=256$,
  $P=256$, $PT=2$, $\Gamma_0=40$, logarithmic annealing scheme).}
\end{figure}

The typical evolution of the quantum annealing in 1d is
illustrated with snap-shots of the spin configurations in Fig.\
\ref{fig2}. The $256$ replicas of the original classical system of
size $L=256$ lie in the horizontal direction. Black pixels represent
spins which match with the orientation in the ground state configuration,
while white pixels correspond to incorrectly aligned spins. Note how the
effectively classical system is strongly correlated along the Trotter
direction.

\subsection{Quantum annealing: one dimension}

The Suzuki-Trotter mapping is exact only in the limit $PT\to\infty$.
This means that the quantum nature of annealing should not be seen when
the value of $PT$ is too small. On the other hand in order to find the
ground state of the system the temperature $T$ needs to be taken to
zero. In the ideal case one should have $PT\to\infty$ and $T=0$,
simultaneously, which is impossible in practice.

\begin{figure}[t]
\includegraphics[width=0.85\columnwidth,angle=-90]{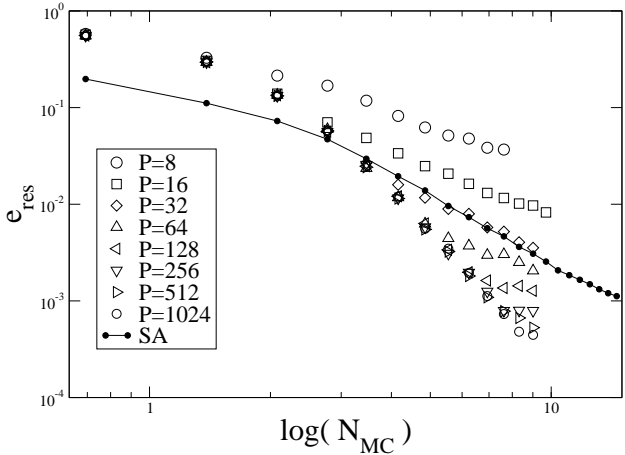}
\caption{\label{dat1d_1} 
  A doubly logarithmic plot of residual energies in 1d with 
  $J=1, L=10^4$, $PT=4$, $P=8...1024$.  For comparison we also show the
  data corresponding to the classical simulated annealing (SA). In
  order to compare the used Monte Carlo time, in the case of QA 
  $N_{MC}$ has to be multiplied with the number of replicas $P$.  
}
\end{figure}

With $P=1$ one simply performs Metropolis dynamics,
and as $P$ is increased the annealing should become more
efficient. This is demonstrated in Fig.\ \ref{dat1d_1} for 1d
systems of $10^4$ spins. The effective temperature is kept
constant $PT=4$ while the number of the Trotter replicas $P$ is varied. 
The results are averaged over $10$ realizations of disorder, each.
   
When $P$ is increased sufficiently ($P>32$) one
can observe a region where the annealing rate of QA is considerably
higher compared to SA. With increasing $P$ this region grows and we
expect that this is the asymptotic behavior for $P\to\infty$. 
Since the Suzuki-Trotter mapping is exact in this limit, one can
assume that this region reflects the properties of quantum
annealing with a finite, non-zero quench rate.
For finite values of $P$ the system falls out of this 
quantum annealing
regime to the classical Metropolis behavior as $N_{MC}$ is
increased. In the classical regime the system is fully correlated in
the Trotter direction and the annealing process cannot anymore take an
advantage of the extra, non-classical dimension. This means that in 
order to maintain a fast annealing rate for large values of $N_{MC}$ 
larger $P$ values are needed as well, as was also observed in 
Ref.~\cite{martonak}. Due to the finite simulation temperature 
the residual energy finally saturates to some non-zero value.

\begin{figure}[t]
\includegraphics[width=0.85\columnwidth,angle=-90]{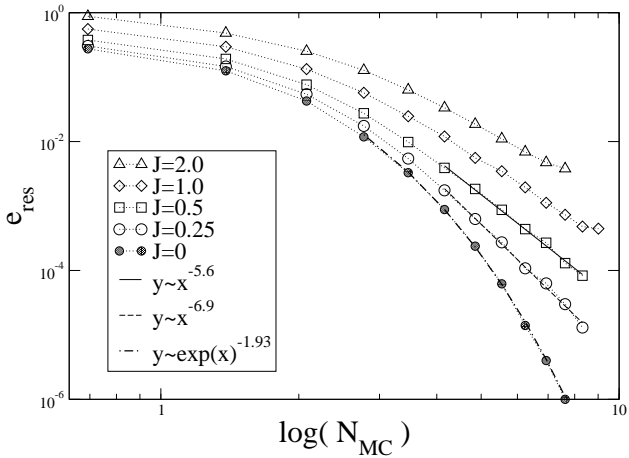}
\caption{\label{dat1d_2} A doubly logarithmic plot of 1d residual
  energies, $L=10^4$, $PT=4$. The  $J=0$ case is a problem of
  $L^d$ independent spins for which $e_{res}\sim
  (N_{MC})^{-\zeta}$ with $\zeta\approx 2$. For $J>0$ 
  $e_{res}$ seems to decay logarithmically. With increasing $J$ the
  RFIM becomes more difficult to anneal.
}
\end{figure}

As indicated in Fig.\ \ref{dat1d_2} the decay rate of $e_{res}$
depends on the actual value of the coupling constant $J$.  
The case with $J=0$ and $\langle h^2\rangle=1$ is a problem of $L^d$
independent spins for which the results depend on the used annealing
schedule \cite{kadowaki}. For the logarithmic and rational schedules
(Eq.~(\ref{g_log})) we find a polynomial decay of the residual energy 
$e_{res}\sim (N_{MC})^{-\tilde{\zeta}}$ with $\tilde{\zeta}\approx 2$
whereas in the case of the linear annealing schedule
$\tilde{\zeta}\approx 1$.
For $J>0$ the quantum annealing goes over to the logarithmic regime
$e_{res}\sim \log(N_{MC})^{-\zeta}$. The numerical values of $\zeta$ 
roughly agrees with the estimate $\zeta_{max}=6$ given by Santoro et
al. \cite{santoro}. 
As $J$ grows, and hence the cluster sizes of the ground state, the
annealing efficiency seems to diminish. 
When no random field is applied the dynamics of the
quantum annealing in the limit $\Gamma\to 0$ for $P>>L$ corresponds
to a case of a strongly anisotropic Ising model. In Ref.\
\cite{ferreira} Ferreira et al. have studied the two dimensional
anisotropic Ising system with $J_x>>k_BT>>J_y$. They found that in the  large
time limit the width of the interfaces perpendicular to $x$--direction 
saturates to some finite value.
We have verified numerically that in same
limit the quantum annealing ends up in a situation where the residual
energy comes from rough, fluctuating boundaries, positioned
perpendicular to the Trotter direction. 

\begin{figure}[t]
\includegraphics[width=0.85\columnwidth,angle=-90]{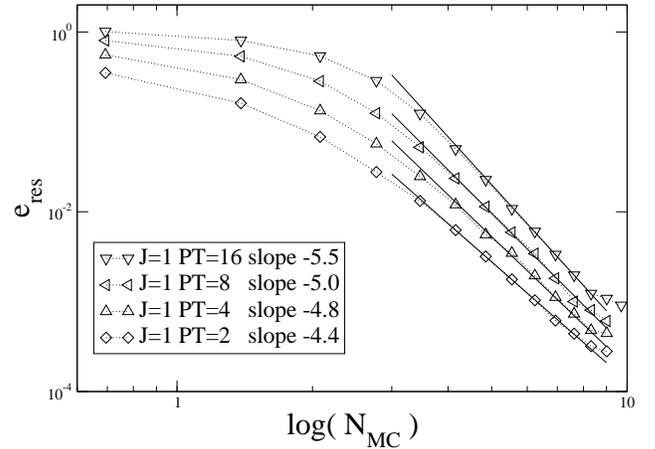}
\caption{\label{dat1d_3} A doubly logarithmic plot of residual
  energies in 1d with $L=10^4$, $J=1$. 
  The annealing rate increases with $PT$ at the cost of growing
  amplitude (prefactor).
}
\end{figure}

The scale of the energy barriers is determined by the value of
$J$. Hence, when $J$ is increased one needs larger $PT$ to overcome
the barriers.  The data in Fig.\ \ref{dat1d_3} shows how the annealing
rate grows as $PT$ is increased.  The value of $PT$ has a two-fold
effect. As indicated in Fig.\ \ref{dat1d_3} large values of $PT$ are
desirable in order to minimize the error in the Suzuki-Trotter mapping
and hence to be able to observe the true quantum annealing behavior.  
On the other hand the $PT$ determines the growth rate of $J_{\perp}$
(see Eq.\ \ref{j_perp}). As $PT$ increases one needs larger a $N_{MC}$ in order
to keep the same annealing rate and hence to reach an equally low value
of $e_{res}$.


\subsection{Quantum annealing: 2d and 3d}

\begin{figure}[t]
\includegraphics[width=0.85\columnwidth,angle=-90]{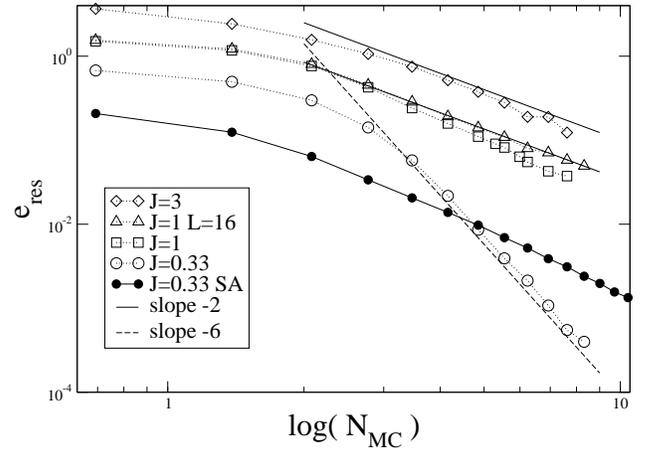}
\caption{\label{dat2d_1} 
  A doubly logarithmic plot of residual energies in 2d,
  $PT=12$, $L=128$, and one data set corresponding to $L=16$. The  
  straight lines are guides for the eye. For  $J\gtrsim 1$ the
  residual energy decays with $\zeta\approx2$. With growing $J$ no
  further decrease of $\zeta$ is observed. For comparison we also show
  the results for classical simulated annealing (SA).}  
\end{figure}

In two dimensions we consider systems with sizes only up to $128\times128$ in
order to be able to consider $P$ up to $1024$.  For low values of $J$,
where the size of the clusters is well below the system size, the QA
performs similarly to the one dimensional case giving $\zeta\approx 6$
for $J=0.33$ (see Fig.\ \ref{dat2d_1} ). With $J$ also the cluster
sizes of the ground state grow reaching the system size 
($128\times128$) approximately at $J=1.5$. From Fig.\ \ref{dat2d_1}
one can see that for $J\gtrsim 1$ one has $\zeta=2$. Whereas in the
case of one dimensional systems with extremely weak disorder 
the value of $\zeta$ could be raised by increasing the effective
temperature $PT$ this seems not to work in two dimensions.
This suggests that there is a fundamental difference in
the performance of QA depending whether the system has a disordered or
ordered ground state. This conclusion is supported by the observation
that with $J=1$ for a larger system ($L=128$, squares in Fig.\
\ref{dat2d_1}) one gets lower residual energies compared to a smaller
system ($L=16$, triangles in Fig.\ \ref{dat2d_1} ) that already
has an ferromagnetically ordered ground state.

\begin{figure}[t!]
\includegraphics[width=0.85\columnwidth,angle=-90]{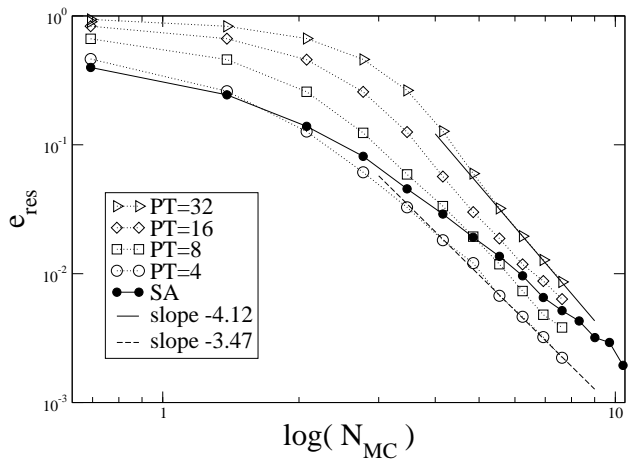}
\caption{\label{dat3d_1} 
  A doubly logarithmic plot of residual energies in 3d, $L=24$,
  $J=0.33$. With increasing $PT$ the effective $\zeta$ is growing. 
  For comparison we also show the results for classical simulated
  annealing (SA).
}
\end{figure}

\begin{figure}[tx]
\includegraphics[width=0.85\columnwidth,angle=-90]{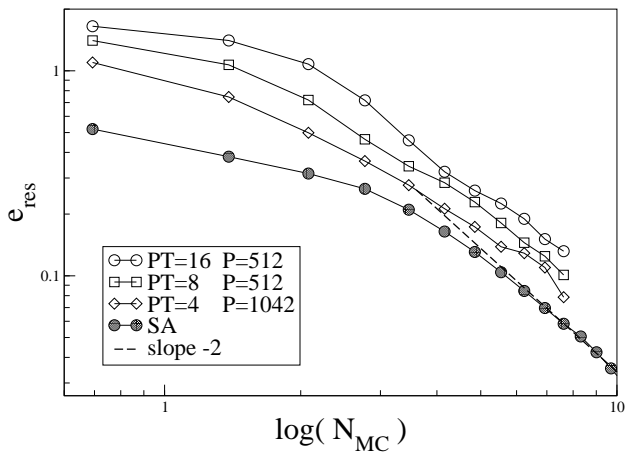}
\caption{\label{dat3d_2} 
  A doubly logarithmic plot of residual energies in 3d, $L=24$,
  $J=0.66$. The decay rate of $e_{res}$ does not vary with $PT$,
  $\zeta\approx 2$. For comparison we also show the results for 
  classical simulated annealing (SA).}  
\end{figure}

As it was already evident in the 2d case also the results for 3d RFIM
show that QA is sensible to the ordering of the underlying system.
For low values of $J$ ($J<J_c\approx 0.44$ \cite{middleton}) where the
system is in the paramagnetic phase also at $T=0$ we find the same
behavior as in one and two dimensions (Fig.\ \ref{dat3d_1}): QA is
faster than SA and $\zeta$ can be tuned towards 6 by increasing
$PT$. Fig.\ \ref{dat3d_2} shows the data corresponding to
$J=0.66$. For the range of parameters that have been used we find that
$\zeta\approx 2$, as in 2d, for a system with an ordered ground state.

\section{Summary}\label{sec:summary}

We have studied numerically quantum annealing in the random field Ising
model and compared our results with classical simulated annealing.
When the system is in the paramagnetic phase we find that
asymptotically QA provides a better decay rate of the residual energy
with $\zeta$ up to $6$ in agreement with the Landau-Zener -picture based
scaling argument presented by Santoro et al. \cite{santoro}. 

We expect that the asymptotic performance of
QA in one and two dimensions does not change by varying the coupling
constant $J$ or the magnetic field strength $h$. 
With growing cluster sizes in the GS one needs increasingly
larger values of $PT$ and $P$ for which the fast annealing rate
with $\zeta\approx6$ could be observed. The requirement of large
values of $P$ in the case of weak random field makes QA from the
practical point of view slower compared to SA. Note the starting point
however: that the RFIM GS can be found effectively with combinatorial
optimization.

In 3d we have presented evidence that the performance of QA depends on
whether the system is in the paramagnetic or ordered phase. Thus, the
situation is actually analogous to the behavior of the SA. In the
paramagnetic phase we find the similar behavior as in one and two
dimensions. In the ordered phase we observe a much slower decay of
$e_{res}$ with $\zeta \approx 2$, so that in fact QA is slower
than SA with a starting temperature $T_0>T_c$.

We conclude with the general observation that the better efficiency of QA 
is most clear when the ground state consists of small clusters, i.e. 
the correlation length of the ground state is short compared to the number 
of the Trotter replicas. Such benefits vanish with an increasing correlation 
length, of the ground state configuration.

\end{document}